\documentclass[twocolumn,showpacs,preprintnumbers, prl,
               superscriptaddress,nofootinbib]{revtex4}
\usepackage{graphicx, fancybox}
\usepackage{amsmath,amssymb,bm}

\begin{document}

\title{
Effects of global charge conservation on 
time evolution of cumulants of conserved charges 
in relativistic heavy ion collisions
}

\author{Miki Sakaida}
\email{sakaida@kern.phys.sci.osaka-u.ac.jp}
\affiliation{
Department of Physics, Osaka University, Toyonaka, Osaka 560-0043, Japan}

\author{Masayuki Asakawa}
\email{yuki@phys.sci.osaka-u.ac.jp}
\affiliation{
Department of Physics, Osaka University, Toyonaka, Osaka 560-0043, Japan}

\author{Masakiyo Kitazawa}
\email{kitazawa@phys.sci.osaka-u.ac.jp}
\affiliation{
Department of Physics, Osaka University, Toyonaka, Osaka 560-0043, Japan}

\begin{abstract}

We investigate the effect of the global charge conservation 
on the cumulants of conserved charges observed in relativistic 
heavy ion collisions in a finite rapidity window, $\Delta\eta$,
with a special emphasis on the time evolution of fluctuations 
in the hadronic medium.
It is argued that the experimental result of the net-electric 
charge fluctuation observed by ALICE 
does not receive effects from
the global charge conservation, because of the finite 
diffusion distance of charged particles in
the hadronic stage.
We emphasize that the magnitude of the effect of the global charge
conservation can be estimated experimentally 
by combining the information on
the $\Delta\eta$ dependences of
various cumulants of conserved charges, 
similarly to
other dynamical properties of the hot medium.

\end{abstract}

\date{\today}

\maketitle

\section{Introduction}

Bulk fluctuations are macroscopic observables, which provide us
various information on microscopic nature of the medium.
In relativistic heavy ion collisions, it is believed that 
the bulk fluctuations observed by event-by-event analysis 
are useful observables, which enable us to characterize 
properties of the hot medium created by collision events 
and to find the QCD critical point
\cite{Koch:2008ia,Stephanov:1998dy,Asakawa:2000wh,Jeon:2000wg,
Ejiri:2005wq,Stephanov:2008qz,Asakawa:2009aj,Friman:2011pf,Kitazawa:2013bta,Nakamura:2013ska,Herold:2014zoa,Nahrgang:2014fza,Morita:2013tu,Fukushima:2014lfa}. 
Recently, experimental investigation of fluctuation observables 
in heavy ion collisions has been actively performed at the 
Relativistic Heavy Ion Collider (RHIC) and the Large Hadron 
Collider (LHC) \cite{STAR,PHENIX,ALICE}. 
Numerical analyses of higher order cumulants in equilibrium have 
also been carried out in lattice QCD Monte Carlo simulations 
\cite{lattice}.

Recently, among the fluctuation observables, especially higher-order 
cumulants of conserved charges
have acquired much attention. 
One of the important properties of the conserved-charge 
fluctuations compared with non-conserving ones is that 
conserved charges can be defined
unambiguously as Noether currents and as a result
their fluctuations can be calculated without ambiguity
in a given theory,
for example, QCD \cite{lattice}.
Moreover, various ways to reveal the medium properties using 
conserved-charge fluctuations, especially non-Gaussianity and 
mixed cumulants, have been suggested \cite{Kitazawa:2013bta}.

Among the conserved-charge fluctuations, 
net baryon number \cite{KA} and net electric charge 
fluctuations are observable in relativistic heavy ion collisions.
One of the important properties of these 
flucatuations is that their higher order 
cumulants normalized by a conserved quantity are 
suppressed in the deconfined medium reflecting the fact that 
the charges carried by elementary excitations 
are smaller in the deconfined medium
\cite{Asakawa:2000wh,Jeon:2000wg,Ejiri:2005wq}.
Recent experimental result on the net electric charge fluctuation 
by the ALICE Collaboration at the LHC \cite{ALICE} shows that the value 
of the second order cumulant of the net electric charge 
$\langle (Q_{\rm(net)})^2 \rangle_c 
= \langle (\delta Q_{\rm(net)})^2 \rangle$
normalized by
the total number of charged particles
$\langle  Q_{\rm(tot)} \rangle$,
\begin{align}
\frac{\langle (Q_{\rm(net)})^2 \rangle_c }{ \langle  Q_{\rm(tot)} \rangle },
\end{align}
with $\delta Q=Q-\langle Q \rangle$ is significantly suppressed 
compared with the one in the equilibrated hadronic medium. 
This result is reasonably understood if one interprets the 
suppression as a survival of the small fluctuation generated in 
the primordial deconfined medium \cite{Asakawa:2000wh,Jeon:2000wg,
Shuryak:2000pd,Aziz:2004qu,Kitazawa:2013bta}.

The experimental result of the ALICE Collaboration also shows that 
the suppression of 
$\langle (Q_{\rm(net)})^2 \rangle_c/\langle  Q_{\rm(tot)} \rangle$ 
becomes more prominent as the pseudo-rapidity window to count the number of particles,
$\Delta\eta$, is taken to be larger.
This $\Delta\eta$ dependence can also be reasonably explained 
with the above interpretation.
This is because the approach of the 
magnitude of the fluctuation
to the equilibrated values of the hadronic medium is slower 
as the volume to count the conserved-charge number 
is taken larger \cite{Shuryak:2000pd,Kitazawa:2013bta}.
In Ref.~\cite{Kitazawa:2013bta}, it is also pointed out that the combined 
experimental information on the $\Delta\eta$ dependences of various 
cumulants of conserved charges enables us to verify the above 
picture on the second-order fluctuation.

However, there exists another mechanism 
called the global charge conservation (GCC)
to cause the suppression
of $\langle(\delta Q_{\rm(net)})^2\rangle$ compared with the 
thermal value. 
If one counts a conserved charge in the total system,
created by the heavy-ion collisions,
there are no event-by-event fluctuations because 
of the charge conservation.
This fact is referred to as the GCC
\cite{Koch:2008ia}.
Owing to the finiteness of the hot medium generated in heavy ion 
collisions, the fluctuations in a finite $\Delta\eta$ range, 
$\langle (\delta Q_{\rm(net)})^2 \rangle_{\Delta\eta}$,
are also affected by the GCC.
Moreover, this effect is more prominent for larger $\Delta\eta$.
In Refs.~\cite{Jeon:2000wg,Bleicher:2000ek}, 
on the assumption that the equilibration is established in 
the final state of the heavy ion collision,
the magnitude of this effect at finite $\Delta\eta$ 
is estimated as 
\begin{align}
\langle (\delta Q_{\rm(net)})^2 \rangle_{\Delta\eta}
= \langle (\delta Q_{\rm(net)})^2 \rangle_{\rm GC}\ (1-\Delta\eta/\eta_{\rm tot}),
\label{eq:bleicher}
\end{align}
where $\langle (\delta Q_{\rm(net)})^2 \rangle_{\rm GC}$ is the 
fluctuation in the grand canonical ensemble and 
$\eta_{\rm tot}$ denotes the total length of the system along the rapidity 
direction. 
Note that the system is assumed to be expanding longitudinally 
with the Bjorken scaling.

It, however, should be noted that 
the suppression of $\langle (\delta Q_{\rm(net)})^2 \rangle / \langle \delta Q_{\rm(tot)} \rangle$ 
observed at ALICE \cite{ALICE}
is more significant than the one
described solely by Eq.~(\ref{eq:bleicher}) 
as we will discuss in Sec.~\ref{sec:resultALICE}.
This result shows that there exists another contribution
for the suppression besides the GCC, such as 
the nonthermal effect
as originally addressed in Refs.~\cite{Asakawa:2000wh,Jeon:2000wg}.
Since Eq.~(\ref{eq:bleicher}) assumes the equilibration,
when the fluctuation is not equilibrated 
the effect of the GCC at ALICE
will be modified from this formula.
The effect of the GCC at LHC energy, therefore, has to be 
revisited with the nonequilibrium effects incorporated.

In the present study, we investigate the effect of the 
GCC on cumulants of conserved charges 
under such nonequilibrium circumstances, by describing the time
evolution of fluctuations in a system with finite volume.
We extend the analyses in 
Refs.~\cite{Shuryak:2000pd,Kitazawa:2013bta} to the case 
with a finite volume with reflecting boundaries.
We also discuss the effects of the GCC
on higher-order cumulants of conserved charges for the 
first time.

By comparing our result with the experimental
result at ALICE in Ref.~\cite{ALICE}, we find that the 
net electric charge fluctuation in the rapidity window observed 
by this experiment is hardly affected by the GCC. 
This result comes from the fact that the rapidity at 
which the fluctuations are affected by the GCC 
is approximately limited within the average
diffusion distance of each particle from the boundaries.
Accordingly, when $\eta_{\rm tot}$ is sufficiently large 
the fluctuations observed at mid-rapidity region is not 
affected by the GCC.
We also argue that the combination of the cumulants of 
conserved charges enables us to confirm this picture
experimentally.

\section{Stochastic formalism to describe diffusion of hadrons}

\subsection{Model}
\label{sec:model}

In the present study, we consider
heavy ion collisions with sufficiently large $\sqrt{s_{\rm NN}}$,
at which the mid-rapidity region has an approximate boost 
invariance.
Useful coordinates to describe such a system are the 
space-time rapidity $\eta$ and proper time $\tau$. 
We denote the net number of a conserved charge per 
unit $\eta$ as $n(\eta,\tau)$, and 
set $\tau=\tau_0$ at hadronization, which phenomenologically takes place at 
almost the same time with chemical freeze-out 
at sufficiently large $\sqrt{s_{\rm NN}}$.

Because of the local charge conservation, 
the probability distribution of $n(\eta,\tau_0)$ at hadronization 
inherits from the one that existed in 
the deconfined medium \cite{Koch:2008ia}. 
After hadronization, particles diffuse 
and rescatter, and the distribution of $n(\eta,\tau)$ 
continues to approach the one of the equilibrated hadronic medium
until kinetic freezeout at $\tau = \tau_{\rm fo}$.
The particle number in a rapidity window $\Delta\eta$ at mid-rapidity 
is given by 
\begin{align}
Q(\Delta\eta,~\tau) = \int_{-\Delta\eta/2}^{\Delta\eta/2} d\eta \ n(\eta,\tau).
\label{eq:Q}
\end{align}
In the following, we investigate the time evolution of the probability distribution
of $Q(\Delta\eta,~\tau)$ in hadronic medium with a finite volume until $\tau = \tau_{\rm fo}$.
We then obtain the cumulants of the conserved charge at 
$\tau = \tau_{\rm fo}$, $\langle Q(\Delta\eta,~\tau_{\rm fo})^n \rangle_c$
as functions of $\Delta\eta$, of which
the comparison with experiments will turn out to enable us to extract 
the diffusion constants of the hadronic medium and initial 
fluctuations at $\tau=\tau_0$, besides the effect of 
the GCC.
Strictly speaking, the experiments measure the particle numbers 
in a {\it pseudo}-rapidity window, while in our study we 
investigate the distribution in the {\it space-time} rapidity.
In the following, we assume the exact Bjorken scaling correspondence
between the space-time rapidity of a particle and its 
{\it momentum-space} rapidity which is almost identical with the 
pseudo rapidity. 
The effects of the violation of the correspondence 
owing to thermal motion of particles and transverse 
expansion will be discussed in a future publication \cite{yohnishi}.

Here, we note that
the second order cumulant of Eq.~(\ref{eq:Q}), $\langle (Q(\Delta\eta))^2 \rangle_c$ 
is directly related to the correlation function, 
$\langle \delta n(\eta_1) \delta n(\eta_2) \rangle$, 
as the latter is obtained by differentiating the former.
The correlation function is further related to 
the balance function \cite{Bass:2000az,Jeon:2001ue,Ling:2013ksb,Pratt:2012dz}.
The experimental information on 
the $\Delta\eta$ dependence of $\langle (Q (\Delta\eta))^2 \rangle_c$, 
therefore, is in principle the same as those obtained from these
functions.
On the other hand, higher order cumulants of $Q$ contain 
information which cannot be described by the two-point 
correlation function.

In this study, 
to describe the time evolution of fluctuation of $Q(\tau)$
we adopt the diffusion master equation \cite{Kitazawa:2013bta}. 
In this model, we divide the system with the total rapidity length 
$\eta_{\rm tot}$ into $M$ discrete cells with an equal finite 
length $a=\eta_{\rm tot}/M$. We then consider a single species 
of particles for the moment, and denote the particle number 
existing in the $m$th cell as $n_m$ and the probability distribution 
that each cell contains $n_m$ particles as $P(\bm{n}, \tau)$ with 
$\bm{n}=(n_{0},n_{1}, \cdots ,n_{m}, \cdots ,n_{M-2},n_{M-1} )$.
The model will be extended to the case of multi-particle species later.
Finally, we assume that each particle moves to the adjacent cells 
with a probability $\gamma (\tau)$ per unit proper time, 
as a result of microscopic interactions. 
The probability $P(\bm{n},\tau)$ then obeys the differential 
equation 
\begin{align}
\partial_\tau& P(\bm{n},\tau)
= \gamma(\tau) \sum_{m=0}^{M-1}[( n_{m} + 1 ) 
\{ P(\bm{n}+\bm{e}_{m}-\bm{e}_{m+1},\tau)\nonumber \\
\quad &+ P(\bm{n} +\bm{e}_{m}-\bm{e}_{m-1},\tau) \} 
- 2 n_m P(\bm{n},\tau) ]\bigr],\!
\label{eq:DME}
\end{align}
where $\bm{e}_{m}$ is a unit vector whose all components are 
zero except for $m$th one, which takes unity. 
In order to take account of the finite size of the hot medium,
we further require that the cells at both ends, at 
$m=0$ and $M-1$, exchange partcles only with inner adjacent cells,
$m=1$ and $M-2$, respectively.
After solving Eq.~(\ref{eq:DME}) exactly, we take the 
continuum limit, $a\to0$.
In this limit, each particle in this model behaves as 
a Brownian particles without correlations with one another 
\cite{Kitazawa:2013bta,Einstein}.

The average and the Gaussian fluctuation of $n(\eta,\tau)$ 
in Eq.~(\ref{eq:DME}) in the continuum limit agree 
with the ones of the stochastic diffusion equation 
\cite{Shuryak:2000pd}
\begin{align}
\partial_\tau n(\eta,\tau) 
= D(\tau) \partial_\eta^2 n(\eta,\tau)+\partial_\eta \xi(\eta,\tau) ,
\label{eq:diffusion}
\end{align}
with two reflecting boundaries
when one sets $D(\tau)=\gamma(\tau)a^2$ \cite{Kitazawa:2013bta}.
Here, $\xi(\eta,\tau)$ is the temporarily-local 
stochastic force, whose property is 
determined by the fluctuation-dissipation relation.
On the other hand, it is known from a general argument
on the Markov process that the fluctuation of
Eq.~(\ref{eq:diffusion}) in equilibrium becomes of Gaussian 
unless $D(\tau)$ explicitly depends on $n(\eta,\tau)$ 
\cite{gardiner,Kitazawa:2013bta}.
This property of Eq.~(\ref{eq:diffusion}) is not suitable 
to describe the higher order cumulants observed in 
relativistic heavy ion collisions, since the experimentally-measured
cumulants \cite{STAR,ALICE} take nonzero values
close to the equilibrated ones. 
On the other hand, the diffusion master equation
Eq.~(\ref{eq:DME}) can give rise to nonzero higher order cumulants 
in equilibrium, because of the discrete nature of the 
particle number \cite{Kitazawa:2013bta}.
This is the reason why we employ the diffusion master equation
instead of the stochastic diffusion equation.

\subsection{Solving diffusion master equation}

Next, we determine the time evolution of cumulants 
by solving Eq.~(\ref{eq:DME}). 
The following numerical procedure is similar to the one 
in Ref.~\cite{Kitazawa:2013bta}, while the introduction of boundaries 
gives rise to a new complexity.
Determination of the initial condition also becomes more
involved owing to the GCC.

We first consider the time evolution of the probability 
$P(\bm{n},\tau)$ with a fixed initial condition
\begin{align}
P(\bm{n},0) = \prod_{m=0}^{M-1} \delta_{n_m,N_m},
\label{eq:fixed}
\end{align}
namely the initial particle numbers are fixed as $n_m(\tau=\tau_0)=N_m$ for all $m$ without fluctuations.
By introducing the factorial generating function, 
\begin{align}
G_{\rm f} ( \bm{s},\tau )
= \sum_{\bm{n}} \prod_{m=0}^{M-1} s_m^{n_m} P( \bm{n},\tau ),
\label{eq:G_f}
\end{align}
 Eq.~(\ref{eq:DME}) is transformed as
\begin{align}
\partial_\tau G_{\rm f} (\bm{s},&\tau)
= \gamma(\tau)\Bigl[(s_1-s_0) \partial_{s_0} 
+(s_{M-2}-s_{M-1}) \partial_{s_{M-1}} 
\nonumber \\
&+\sum_{m=1}^{M-2} ( s_{m+1} - 2s_m + s_{m-1} ) \partial_{s_m} 
\Bigr] G_{\rm f} (\bm{s},\tau).
\label{eq:G_f:ME}
\end{align}
 Solving Eq.~(\ref{eq:G_f:ME}) with the method of characteristics,
one obtains the solution with the initial condition 
Eq.~(\ref{eq:fixed}) as 
\begin{align}
G_{\rm f} ( \bm{s}, \tau )
= \prod_{m=0}^{M-1} \left( \sum_{k=0}^{M-1} r_k v_{km} 
e^{-\Omega_k(\tau)} \right)^{N_m} ,
\label{eq:G_f(t)}
\end{align}
where
\begin{align}
r_k &= \sum_{m=0}^{M-1} s_{m}u_{mk},
\label{eq:FT:ts}
\end{align}
with
\begin{align}
u_{mk}&=\frac1M  \cos \frac{\pi k(m+1/2)}M ,
\end{align}
is the Fourier transform of $s_m$ and
\begin{align}
\Omega_k(\tau) &= \int_{\tau_0}^{\tau}d\tau' 
\gamma(\tau')\left(\frac{\pi k}M \right) ^2 .
\end{align}
The inverse Fourier transform of Eq.~(\ref{eq:FT:ts}) is given by 
$s_m = \sum_{k=0}^{M-1} r_k v_{km}$ with 
\begin{align}
v_{km}&=\left\{ \begin{array}{ll}
1&(k=0)\\
2\cos \displaystyle{\frac{\pi k(m+1/2)}M}&(k\neq0)\\
\end{array}. \right.
\label{eq:FT:v}
\end{align}

The cumulants of $n_m$ are given by
\begin{align}
\langle n_{m_1} n_{m_2} \cdots n_{m_l} \rangle_{\rm c}
=& \left. \frac{\partial^l  K }
{ \partial \theta_1 \cdots \partial \theta_l } \right|_{\bm{\theta}=0} ,
\label{eq:nm}
\end{align}
with $K(\bm{\theta},\tau)= \log G_{\rm f} ( \bm{s}, \tau )|_{s_m=e^{\theta_m}}$.
For example, the first and second order cumulants are calculated to be
\begin{align}
\langle n_m \rangle_{\rm c}
=& \sum_{m'=0}^{M-1}\sum_{k=0}^{M-1} N_{m'}  u_{mk} v_{km'} e^{-\Omega_k(\tau)} ,
\label{eq:1}
\\
\langle {n}_{m_1} {n}_{m_2} \rangle_{\rm c}=& \delta_{m_1 m_2}\langle n_{m_1} \rangle_{\rm c}-\sum_{m'=0}^{M-1}\sum_{k_1,k_2}^{M-1}N_{m'}u_{m_1k_1}u_{m_2k_2}\nonumber\\
&\times v_{k_1m'}v_{k_2m'}e^{-\left[\Omega_{k_1} (\tau)+\Omega_{k_2} (\tau)\right]},
\label{eq:2}
\end{align}
respectively.

Next, we take the continuum limit $a\to0$.
We set the boundaries at $\eta=\pm\eta_{\rm tot}/2$.
Then, the lower space-time rapidity side of $m$th cell 
is located at $\eta=(m-M/2)a$. 
The particle number per unit rapidity is $n(\eta)=n_m/a$. The probability distribution
$P(\bm{n},\tau)$ in Eq.~({\ref{eq:DME}) becomes a functional of the particle number density $n(\eta)$, 
which is denoted as $P[n(\eta),\tau]$. 
From Eq.~(\ref{eq:1}), one finds that the average of $n(\eta)$
coincides with the solution of Eq.~(\ref{eq:diffusion})
with $D(\tau)=\gamma(\tau)a^2 $.
We thus take the continuum limit with fixed $D(\tau)$.
Using Eq.~(\ref{eq:2}), it is also confirmed that the Gaussian 
fluctuation in our model agrees with that of Eq.~(\ref{eq:diffusion}) 
for sufficiently smooth initial condition in this limit.

With the fixed initial condition $n(\eta,\tau_0)=N(\eta)$, 
the cumulants of Eq.~(\ref{eq:Q}) at proper time $\tau$ 
are calculated to be
\begin{align}
\langle (Q(\Delta\eta,~\tau))^n \rangle_{\rm c} 
= \int_{-\eta_{\rm tot}/2}^{\eta_{\rm tot}/2} d\eta \ N(\eta) H_n(\eta) ,
\label{eq:<Q^n>}
\end{align}
where
\begin{align}
H_1(\eta) =& I(\eta) ,
\\
H_2(\eta) =& I(\eta) - I(\eta)^2 ,
\\
H_3(\eta) =& I(\eta) - 3 I(\eta)^2 + 2 I(\eta)^3 ,
\\
H_4(\eta) =& 
I(\eta) - 7 I(\eta)^2 + 12 I(\eta)^3- 6 I(\eta)^4 ,
\end{align}
with
\begin{align}
I(\eta)&=\frac{\Delta\eta}{\eta_{\rm tot}}\sum_{k=-\infty}^{\infty} \cos \left( \frac{\pi k \eta}{\eta_{\rm tot}}\right) \frac{ \sin \left({\displaystyle \frac{\pi k\Delta\eta}{2\eta_{\rm tot}} }\right)}{\left( {\displaystyle \frac{\pi k\Delta\eta}{2\eta_{\rm tot}} }\right)} \nonumber\\ &\times \cos \left(\frac{\pi k}{2}\right) \exp\left[-\frac{1}{2}\left(\frac{\pi k d(\tau)}{\eta_{\rm tot}}\right)^2\right].
\label{eq:I_X}
\end{align}
Here, 
\begin{align}
d(\tau)=\left[2\int_{\tau_0}^{\tau}d\tau'D(\tau')\right]^{1/2}
\label{eq:d(tau)}
\end{align}
is the average diffusion length of each Brownian particle \cite{Einstein}.

Next, we extend the above results to the 
cases with general initial conditions with non-vanishing initial fluctuations. 
In the following, we also extend the formula to treat
cumulants of the difference of densities of two particle species, $n_1(\eta,\tau)$ and $n_2(\eta,\tau)$,
\begin{align}
Q_{\rm (net)}(\Delta\eta,~\tau)=\int_{-\Delta\eta/2}^{\Delta\eta/2} d\eta \ (n_1(\eta,\tau)-n_2(\eta,\tau)),
\label{eq:dif}
\end{align}
in order to consider cumulants of conserved charges, which are given by the difference of particle numbers;
in the following we assume that the net number Eq.~(\ref{eq:dif}) is 
a conserved charge.
If there exists the initial fluctuation $P[N_1(\eta),N_2(\eta),\tau_0]=J[N_1(\eta),N_2(\eta)]$,
the probability distribution of a conserved charge $P[n_1(\eta),n_2(\eta),\tau]$ is given by the superposition of the solutions of fixed initial condition
\begin{align}
P[n_1,n_2,\tau]=\sum_{\{N_1,N_2\}}J[N_1,N_2]P_{N_1}[n_1,\tau]P_{N_2}[n_2,\tau],
\label{eq:gene}
\end{align}
where $P_N[n,\tau]$ is the solution of Eq.~(\ref{eq:DME}) with 
the fixed initial condition $n(\eta,\tau_0)=N(\eta)$ and the sum runs over functional space of $N_1(\eta)$ and $N_2(\eta)$.
Using Eq.~(\ref{eq:gene}) and the same technique
used in Ref.~\cite{Kitazawa:2013bta},
one can obtain the cumulants of Eq.~(\ref{eq:dif}).

\subsection{Initial condition}

Next, let us constrain the initial condition.
Because we consider a finite system, 
the initial condition at $\tau=\tau_0$
should be determined in accordance with the GCC,
i.e., the fluctuation of the net particle number in 
the total system should vanish in the initial condition.
In the following, we constrain ourselves to the 
initial condition satisfying boost invariance between the two boundaries.
In order to obtain the initial conditions 
which conform the GCC,
here we model the initial configuration by
an equilibrated free classical gas in a finite volume.
In this system,
using the fact that each particle can be observed 
at any spatial points in the system with the same probability, 
one can obtain the correlation functions by taking the 
continuum limit of the multinomial distribution.
The result for
the correlation functions of 
$N_{\rm(net),(tot)}(\eta) = N_1(\eta)\mp N_2(\eta)$
up to the third order
is given by 
\begin{align}
\langle N_i(\eta)\rangle _{\rm c}&=[N_i]_{\rm c},
\label{eq:inicor1}
\\
\langle N_{i_1}(\eta_1) N_{i_2}(\eta_2)\rangle_{\rm c}&=[N_{i_1}N_{i_2}]_{\rm c}(\delta(1,2)-1/\eta_{\rm tot}),
\label{eq:inicor2}
\end{align}
\begin{align}
\lefteqn{\langle N_{i_1}(\eta_1) N_{i_2}(\eta_2) N_{i_3}(\eta_3)\rangle_{\rm c}}
\nonumber \\
&=[N_{i_1}N_{i_2}N_{i_3}]_{\rm c}
(\delta(1,2)\delta(2,3) \nonumber\\
&-[\delta(1,2)+\delta(2,3)+\delta(3,1)]/\eta_{\rm tot} + 2(1/\eta_{\rm tot})^2),
\label{eq:inicor3}
\end{align}
when at least one of $i_n$ is (net)
where the subscript $i_n$ denotes (net) or (tot).
Here, $[N_{i_1}\cdots N_{i_l}]_{\rm c}$ are susceptibilities of the
initial condition in the grand canonical ensemble
\cite{Kitazawa:2013bta}, and 
$\delta(j,k)=\delta(\eta_j-\eta_k)$.
The expression for fourth and higher order terms are 
lengthy but obtained straightforwardly.
The total number is not conserving and can have 
event-by-event fluctuation even when the total system is observed.
Reflecting this property, we also assume that 
the number of $N_{\rm(tot)}$ at $\tau=\tau_0$ in the total system 
obeys the one given by the grand canonical ensemble.
One then finds that when all $i_n$ are 
(tot) in Eqs.~(\ref{eq:inicor1}) - (\ref{eq:inicor3})
terms containing $\eta_{\rm tot}$ do not appear.
We note that Eq.~(\ref{eq:inicor2}) reproduces
Eq.~(\ref{eq:bleicher}).

Using these initial conditions and Eq.~(\ref{eq:<Q^n>}),
one obtains the first four cumulants of 
$Q_{\rm(net)}(\Delta\eta,~\tau)$ as
\begin{widetext}
\begin{align}
\langle Q_{{\rm(net)}} (\Delta\eta) \rangle_{\rm c} 
=&[N_{\rm(net)}]_{\rm c}\Delta\eta,
\label{eq:<Q^1>t}
\\
\langle (Q_{{\rm(net)}} (\Delta\eta))^2 \rangle_{\rm c} 
=& [N_{\rm(tot)}]_{\rm c}[\Delta\eta-F_2(\Delta\eta)]
-[N_{\rm(net)}^2]_{\rm c}[\Delta\eta p-F_2(\Delta\eta)] ,
\label{eq:<Q^2>t}
\\
\langle (Q_{{\rm(net)}} (\Delta\eta))^3 \rangle_{\rm c} 
=&[N_{\rm(net)}]_{\rm c}[\Delta\eta-3F_2(\Delta\eta)+2F_3(\Delta\eta)]-3[N_{\rm(net)}N_{\rm(tot)}]_{\rm c}[\Delta\eta p-\left(1+p \right)F_2(\Delta\eta)+F_3(\Delta\eta)]\nonumber\\
+&[N_{\rm(net)}^3]_{\rm c}[2\Delta\eta p^2-3pF_2(\Delta\eta)+F_3(\Delta\eta)],
\label{eq:<Q^3>t}
\\
\langle (Q_{{\rm(net)}} (\Delta\eta))^4 \rangle_{\rm c} 
=&[N_{\rm(tot)}]_{\rm c}[\Delta\eta-7F_2(\Delta\eta)+12F_3(\Delta\eta)-6F_4(\Delta\eta)]
+3[N_{\rm(tot)}^2]_{\rm c}[F_2(\Delta\eta)-2F_3(\Delta\eta)+F_4(\Delta\eta)]
\nonumber\\
-4&[N_{\rm(net)}^2]_{\rm c}[ \Delta\eta p-(1+3p)F_2(\Delta\eta)+(3+2p)F_3(\Delta\eta)-2F_4(\Delta\eta) ]
\nonumber\\
+&[N_{\rm(net)}^2N_{\rm(tot)}]_{\rm c}\left[2\Delta\eta p^2-p\left(3+2p-\frac{1}{\Delta\eta}F_2(\Delta\eta)\right)F_2(\Delta\eta) +\left(1+2p\right)F_3(\Delta\eta)-F_4(\Delta\eta)\right]\nonumber\\
-&[N_{\rm(net)}^4]_{\rm c}\left[6\Delta\eta p^3-4p\left(3p-\frac{1}{\Delta\eta}F_2(\Delta\eta)\right)F_2(\Delta\eta)+3pF_3(\Delta\eta)-F_4(\Delta\eta)\right],
\label{eq:<Q^4>t}
\end{align}
\end{widetext}
with 
\begin{align}
F_n (\Delta\eta)= \int_{-\eta_{\rm tot}/2}^{\eta_{\rm tot}/2} d\eta [I(\eta)]^n,
\label{eq:F^n}
\end{align}
and $p=\Delta\eta/\eta_{\rm tot}$.
Using $\lim_{\tau\to\tau_0}F_n(\Delta\eta)=\Delta\eta$, one can check
that only the first terms of Eqs.~(\ref{eq:<Q^1>t}) - (\ref{eq:<Q^4>t}) 
survive at $\tau=\tau_0$.
The initial condition Eqs.~(\ref{eq:inicor1}) - (\ref{eq:inicor3})
thus are reproduced for $\tau=\tau_0$.

\subsection{Equilibration}

If $D(\tau)$ is constant, although physically this is not the case,
the system settles down to an equilibrium state
in the large $\tau$ limit.
The cumulants in this limit is obtained by substituting
$\lim_{\tau\to\infty}F_n=\eta_{\rm tot}p^n$ for $n\ge2$ to
Eqs.~(\ref{eq:<Q^1>t}) - (\ref{eq:<Q^4>t}).
One then obtains for second and third order 
\begin{align}
\langle (Q_{{\rm(net)}} (\Delta\eta))^2 \rangle_{\rm c} &= 
[N_{\rm(tot)}]_{\rm c}(1-p),
\label{eq:Qp2}
\\
\langle (Q_{{\rm(net)}} (\Delta\eta))^3 \rangle_{\rm c} &= 
[N_{\rm(net)}]_{\rm c}(1-p)(1-2p),
\label{eq:Qp3}
\end{align}
respectively, with $p=\Delta\eta/\eta_{\rm tot}$.
These $p$ dependences are consistent with the binomial
distribution functions \cite{KA}; in particular, Eq.~(\ref{eq:Qp2})
is nothing other than
Eq.~(\ref{eq:bleicher}).
Equation~(\ref{eq:Qp3}) is the generalization of 
Eq.~(\ref{eq:bleicher}) to the third order.
On the other hand, the $p$ dependence of the fourth order cumulant 
in the $\tau\to\infty$ limit does not agree with the binomial form,
\begin{align}
\langle (Q_{{\rm(net)}}  (\Delta\eta))^4 \rangle_{\rm c} \propto (1-p)(1-6p+6p^2),
\label{eq:Qp4}
\end{align}
if the initial fluctuation of the total number, 
$[N_{\rm (tot)}^2]_{\rm c}$ has 
a nonzero value.

\section{Effects of the GCC}

Next, let us study how the GCC affects
the rapidity window dependence of the cumulants of conserved 
charges.
Because the odd order cumulants are difficult to measure at LHC energy 
owing to their smallness, in this study 
we limit our attention only to the second and fourth order cumulants.

The cumulants of conserved charges Eqs.~(\ref{eq:<Q^1>t}) -
(\ref{eq:<Q^4>t}) 
are described by using three variables
$\Delta\eta$, $\eta_{\rm tot}$, and $d(\tau)$,
having the dimension of space-time rapidity.
Among them, the diffusion length $d(\tau)$ is an increasing 
function of the elapsed time $\tau-\tau_0$.
When we describe the time evolution in what follows, we use 
the dimensionless parameters
\begin{align}
L \equiv \eta_{\rm tot}/d(\tau),\quad T \equiv d(\tau)/\eta_{\rm tot}.
\label{eq:para}
\end{align}

\subsection{Without initial fluctuation}
\label{sec:result0}

\begin{figure}
  \begin{center}
          \includegraphics[keepaspectratio, angle=-90, clip, width=9cm]{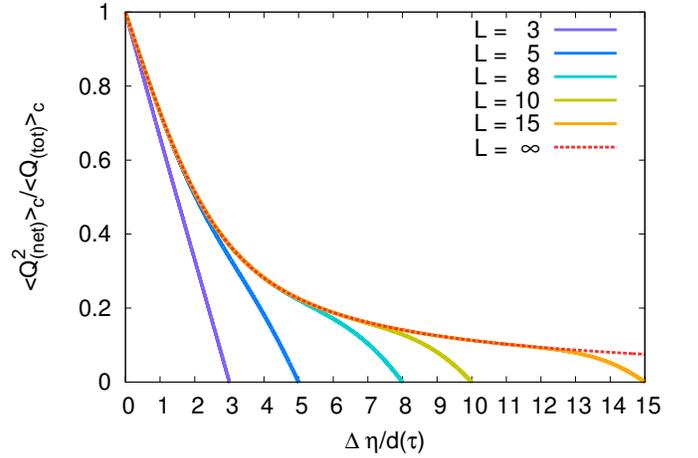}
\caption{Second order cumulants of conserved charges without initial fluctuation as a function of $\Delta\eta/d(\tau)$
with five finite values of the parameter $L$ and infinite $L$.}
       \label{fig:21b0}
        \end{center}
\end{figure}

First, we consider the $\Delta\eta$ dependence for 
the fixed initial condition,
i.e., all fluctuations vanish at $\tau=\tau_0$,
\begin{align}
[N_{{\rm(net)}}^2]_{\rm c}=[N_{{\rm(net)}}^4]_{\rm c}=[N_{{\rm(net)}}^2N_{{\rm(tot)}}]_{\rm c}=[N_{{\rm(tot)}}^2]_{\rm c}=0.
\label{eq:ini1}
\end{align}
In Fig.~\ref{fig:21b0}, we show the $\Delta\eta$ dependence of 
$\langle (Q_{\rm(net)}  (\Delta\eta))^2 \rangle_{\rm c}$, for several values of $L$.
The horizontal axis is normalized by $d(\tau)$, while the vertical
one is normalized by the equilibrated value in an infinite volume, 
$\langle Q_{{\rm (tot)}}  (\Delta\eta) \rangle_{\rm c}$.
For comparison, we also show the result with an infinite volume 
by the dashed and dotted lines.
The figure shows that $\langle (Q_{{\rm (net)}}  (\Delta\eta))^2 \rangle_{\rm c}$ 
vanishes at $\Delta\eta/d(\tau)=L$ for each $L$, namely 
$\Delta\eta=\eta_{\rm tot}$, which is a trivial consequence of 
the GCC.
On the other hand, as $\Delta\eta$ becomes smaller from this value,
the result with finite $L$ approaches the one with infinite 
volume. The effect of $L$ vanishes almost completely 
except for the range 
\begin{align}
\Delta\eta/d(\tau)\gtrsim L-2.
\label{eq:eta-d}
\end{align}
This is somewhat an unexpected result compared with the 
previous estimate Eq.~(\ref{eq:bleicher}).

One can, however, give a physical interpretation to this result
as follows.
We first note that Eq.~(\ref{eq:eta-d}) is rewritten as
\begin{align}
\frac{\eta_{\rm tot}- \Delta \eta}{2} \lesssim d(\tau).
\label{eq:effec}
\end{align}
In this expression, the left-hand side is the distance between 
the left (right) boundary and the left (right) edge of the rapidity window, while the 
right-hand side is the diffusion length of each Brownian particle.
When Eq.~(\ref{eq:effec}) is satisfied, particles which are reflected
by one of the boundaries at least once can enter the rapidity window.
Therefore, the existence of the boundaries can 
affect the fluctuations of conserved charges in the rapidity window.
On the other hand, when the condition Eq.~(\ref{eq:effec}) 
is not satisfied, particles inside the rapidity window do not 
know the existence of the boundaries, in other words, 
the fact that the system is finite.
In the latter
case, therefore, the fluctuations in the rapidity window
are free from the effect of the GCC.

Figure~\ref{fig:21b0} also shows that for $L=3$ 
the $\Delta\eta$ dependence of $\langle (Q_{\rm(net)} (\Delta\eta))^2 \rangle_{\rm c}$ 
becomes almost a linear function; note that this is the behavior
consistent with Eq.~(\ref{eq:bleicher}).
For $L=3$, the diffusion length is almost comparable with the
system size.
When the condition $d(\tau) \gtrsim \eta_{\rm tot}/2$ is satisfied, 
each particle can be anywhere in the system with almost an 
equal probability irrespective of its initial position at $\tau=\tau_0$.
Because this is nothing but the condition for the 
establishment of the equilibration of the system,
the estimate Eq.~(\ref{eq:bleicher}), which relies on the 
equilibration of the system, becomes applicable. 

In this analysis, it is assumed that the particle current vanishes
at the two boundaries. This assumption would not be suitable to 
describe the hot medium created by heavy ion collisions, 
since the hot medium does not have such hard boundaries but
only baryon rich regions.
From the above discussion, however, it is obvious that
the effect of boundaries does not affect the fluctuations in 
the rapidity window unless the condition Eq.~(\ref{eq:effec}) 
is realized irrespective of the types of the boundaries.

\begin{figure}
        \begin{center}       
          \includegraphics[keepaspectratio, angle=-90, clip, width=9cm]{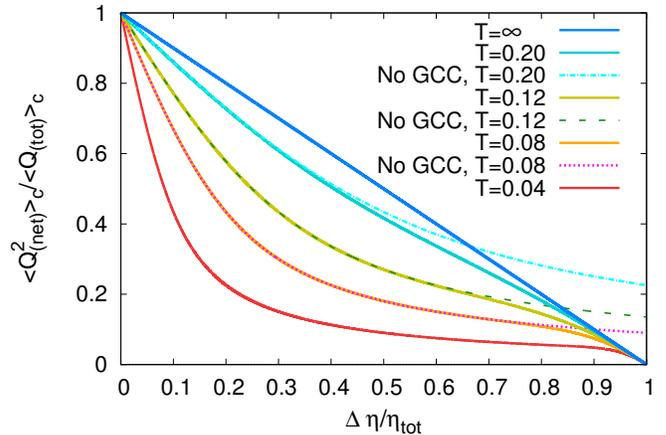}
\caption{
Second order cumulants of conserved charges without initial 
fluctuation as a function of $\Delta\eta/\eta_{\rm tot}$ 
with five values of the parameter $T$.
The corresponding results in an infinite volume are also plotted
for several values of $T$.
}
             \label{fig:22b0}
      \end{center}
\end{figure}
\begin{figure}
      \begin{center}       
          \includegraphics[keepaspectratio, angle=-90, clip, width=9cm]{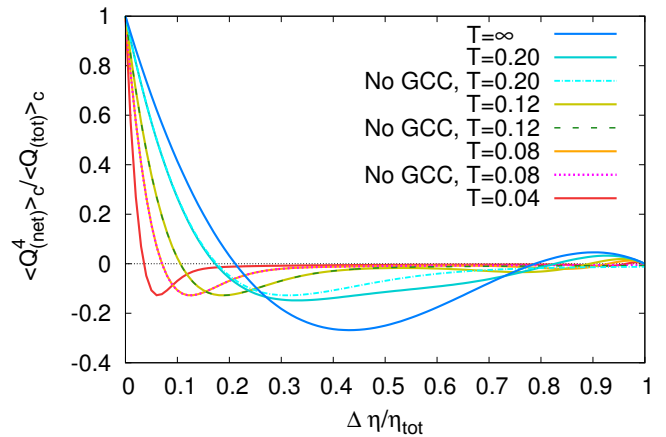}
\caption{Fourth order cumulants of conserved charges without initial fluctuation as a function of $\Delta\eta/\eta_{\rm tot}$ with five values of $T$.}
             \label{fig:4th20}  
      \end{center}
\end{figure}

In Figs.~\ref{fig:22b0} and \ref{fig:4th20}, 
we show $\langle (Q_{{\rm (net)}}  (\Delta\eta))^n \rangle_{\rm c}
/\langle Q_{{\rm (tot)}}  (\Delta\eta) \rangle_{\rm c}$ for $n=2$ and $4$,
respectively, as a function of $\Delta\eta/\eta_{\rm tot}$
with the total length $\eta_{\rm tot}$ for five values of $T$ including infinity. 
When one can estimate the value of $\eta_{\rm tot}$ in experiments
the plots in Figs.~\ref{fig:22b0} and \ref{fig:4th20} 
are suitable for comparison with the experiments
\cite{ALICE}. 
The corresponding results in an infinite volume 
are also plotted for several values of $T$ by
the dashed and dotted lines.
With fixed $\eta_{\rm tot}$, larger $T$ corresponds to larger $\tau$.
Figure~\ref{fig:22b0} shows that as $T$ becomes larger 
the fluctuation increases and approaches a linear function 
representing the equilibration, 
Eq.~(\ref{eq:bleicher}) or (\ref{eq:Qp2}).
Moreover, the comparison of each result with the infinite-volume ones
shows that the GCC can affect 
the fluctuations only for cases with large $\Delta\eta/\eta_{\rm tot}$.
As we shall see later, the largest rapidity windows covered by 
STAR at the top RHIC energy and ALICE correspond to 
$\Delta\eta/\eta_{\rm tot}\simeq0.2$.
Figure~\ref{fig:22b0} suggests that for this rapidity coverage,
when the value of $\langle (Q_{{\rm (net)}} (\Delta\eta))^2 \rangle_{\rm c} $ 
shows a suppression compared with Eq.~(\ref{eq:bleicher}),
the effect of the GCC is negligible.

From Fig.~\ref{fig:4th20}, which shows the 
fourth order cumulants for several values of $T$,
one obtains completely the same conclusion on the effect of
the GCC. The figure shows that the effect of the GCC
is visible only for large values of $\Delta\eta/\eta_{\rm tot}$.
We also notice that in Fig.~\ref{fig:4th20} the 
$\Delta\eta/\eta_{\rm tot}$ dependence in the large $\tau$ limit 
becomes of binomial form, Eq.~(\ref{eq:Qp4}).
This limiting behavior, however, is not realized for 
initial conditions with $[N_{{\rm(tot)}}^2]_{\rm c}\ne0$.

\subsection{Effect of initial fluctuation}

\begin{figure}
\begin{center}
  \includegraphics[keepaspectratio, angle=-90, clip, width=9cm]{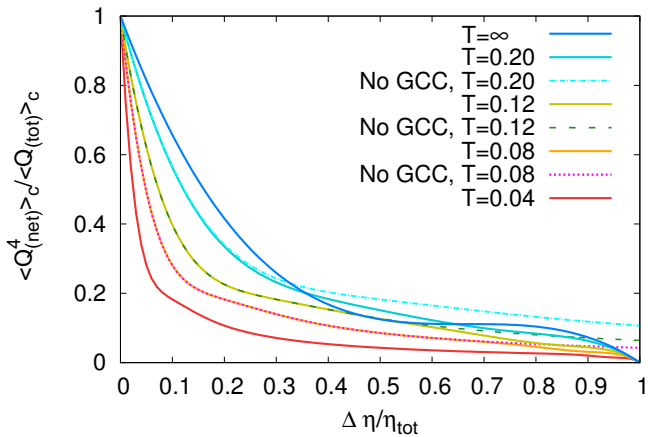}

  \caption{
Fourth order cumulants of conserved charges 
under the initial condition (\ref{eq:incon2}) 
with five values of $T$.}
\label{fig:40001}
\end{center}
\end{figure}

\begin{figure}
\begin{center}
  \includegraphics[keepaspectratio, angle=-90, clip, width=8.5cm]{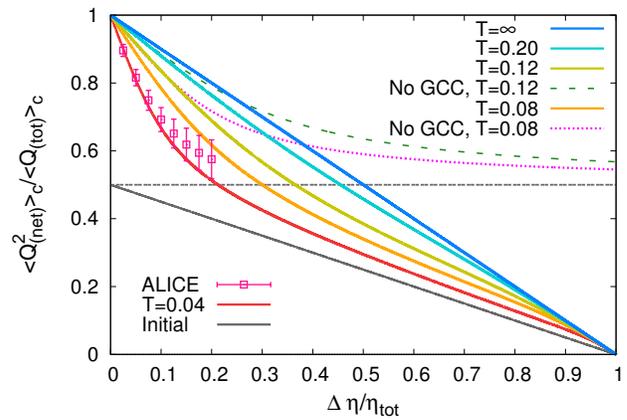}
  \caption{
Second order cumulants of conserved charges under the initial condition 
(\ref{eq:incon3}) with five values of $T$.
The experimental results in Ref.~\cite{ALICE} with $\eta_{\rm tot}=8$
is also plotted by squares for comparison; 
see, Sec.~\ref{sec:resultALICE} for a detailed discussion.
}
\label{fig:2b05}
\end{center}
\end{figure}

\begin{figure}
\begin{center}
  \includegraphics[keepaspectratio, angle=-90, clip, width=9cm]{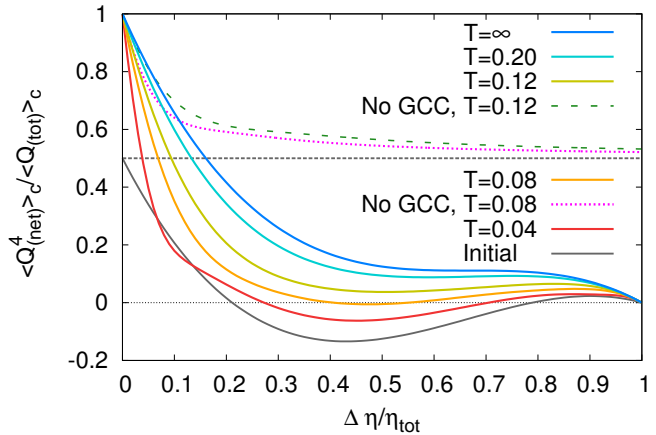}
  \caption{Fourth order cumulants of conserved charges under the initial condition (\ref{eq:incon3}) with five values of $T$.}
  \label{fig:40505051}
\end{center}
\end{figure}

Second, let us look at the $\Delta\eta$ dependence of cumulants 
with initial conditions having nonvanishing fluctuations.
We first consider the effect of the fluctuations of total 
charge number $[N_{\rm(tot)}^2]_{\rm c}$, which is an observable
proposed in Ref.~\cite{Kitazawa:2013bta} as a new probe for
hadronization mechanism.
To see the effect of $[N_{\rm(tot)}^2]_{\rm c}$, we 
set all the cumulants including net-charge at $\tau=\tau_0$ 
zero, $[N_{{\rm(net)}}^2]_{\rm c}=[N_{{\rm(net)}}^4]_{\rm c}
=[N_{{\rm(net)}}^2N_{{\rm(tot)}}]_{\rm c}=0$.
The second-order cumulant does not change from
the previous result in Fig.~\ref{fig:22b0} by including
 $[N_{\rm(tot)}^2]_{\rm c}$, but the fourth-order one does.
In Fig.~\ref{fig:40001}, we show the $\Delta\eta$ dependence of 
the fourth-order cumulant with
\begin{align}
[N_{\rm(tot)}^2]_{\rm c}=[N_{{\rm(tot)}}]_{\rm c},
\label{eq:incon2}
\end{align}
which is realized in the Poissonian case.
By comparing the results with the ones of the infinite volume
shown by the dashed and dotted lines, one obtains the same 
conclusion on the effect of the GCC as in the previous
subsection, i.e., the effect 
alters the fluctuation only for large $\Delta\eta/\eta_{\rm tot}$.

Figure~\ref{fig:40001} also shows that the $\Delta\eta$ 
dependence in Fig.~\ref{fig:40001} is qualitatively different 
from the one in Fig.~\ref{fig:4th20}.
For example, although $\langle (Q_{\rm (net)}(\Delta\eta))^4\rangle_c $ is 
nonmonotonic and becomes negative in Fig.~\ref{fig:4th20},
such behaviors are not observed in Fig.~\ref{fig:40001},
irrespective of the value of $T$.
Moreover, these differences are observed even for 
small $\Delta\eta/\eta_{\rm tot}$.
This result indicates that the magnitude of 
$[N_{\rm(tot)}^2]_{\rm c}$ can be experimentally estimated by 
measuring the $\Delta\eta$ dependence of 
$\langle (Q_{\rm (net)}(\Delta\eta))^4\rangle_c   $ \cite{Kitazawa:2013bta}.

Next, to see the effect of initial fluctuations of conserved charges,
in Figs.~\ref{fig:2b05} and \ref{fig:40505051} 
we show the second and fourth order cumulants with the initial condition 
\begin{align}
[N_{{\rm(net)}}^2]_{\rm c}&=[N_{{\rm(net)}}^4]_{\rm c}=[N_{{\rm(net)}}^2N_{{\rm(tot)}}]_{\rm c}=0.5[N_{{\rm(tot)}}],\nonumber\\
[N_{{\rm(tot)}}^2]_{\rm c}&=[N_{{\rm(tot)}}]_{\rm c}.
\label{eq:incon3}
\end{align}
The value $[N_{{\rm(net)}}^2]_{\rm c}=0.5[N_{{\rm(tot)}}]$ is 
taken from the estimate in Ref.~\cite{Asakawa:2000wh}.
By comparing the results with and without boundaries in these figures, 
one finds that the effect of GCC
is observed already from small $\Delta\eta/\eta_{\rm tot}$ 
in this case.
This is because the initial condition determined in Eqs.~(\ref{eq:inicor2})
and (\ref{eq:inicor3}) already include the effect of the
GCC to some extent.
These initial conditions are determined under an assumption that
the medium just before the hadronization 
consists of equilibriated quarks.
The effect of the GCC on the diffusion in the 
hadronic medium with small $\Delta\eta/\eta_{\rm tot}$ 
is still almost invisible even in this case.

\subsection{Comparison with experimental result at ALICE}
\label{sec:resultALICE}

We finally inspect the $\Delta\eta$ dependence
of the net electric charge fluctuations observed at ALICE \cite{ALICE}
in more detail.
In order to estimate the effect of the GCC, 
we must first determine the magnitude of $\eta_{\rm tot}$.
From the pseudo-rapidity dependence of charged-particle yield 
at LHC energy, $\sqrt{s_{\rm NN}}=2.76$ TeV, in Ref.~\cite{Abbas:2013bpa}, 
we take the value $\eta_{\rm tot}=8$ in the following.
Note that this value is considerably smaller than 
twice the beam rapidity $2y_{\rm beam} \simeq 16$.
While the choice of $\eta_{\rm tot}$ is ambiguous \cite{Abbas:2013bpa},
the following discussion is not altered qualitatively by the choice of 
$\eta_{\rm tot}$ in the range $8\lesssim \eta_{\rm tot} \lesssim 12$.
With the choice $\eta_{\rm tot}=8$, 
the maximum rapidity coverage of the $2\pi$ detector, TPC, of ALICE, 
$\Delta\eta=1.6$ \cite{ALICE}, corresponds to
$\Delta\eta/\eta_{\rm tot} = 0.2$.

In Fig.~\ref{fig:2b05}, we overlay the experimental result
of $\langle (Q_{\rm (net)} (\Delta\eta))^2 \rangle_c/\langle Q_{\rm (tot)} (\Delta\eta) \rangle = D/4$
in Fig.~3 of Ref.~\cite{ALICE} for the centrality bin $0-5\%$
with $\eta_{\rm tot}=8$.
From the figure, 
one can immediately conclude that 
the suppression of $\langle (Q_{\rm (net)} (\Delta\eta))^2 \rangle_c$ in 
this experiment cannot be explained solely by the naive formula 
of the GCC, Eq.~(\ref{eq:bleicher}).
From the discussion in the previous subsections,
it is also concluded that the effect of the GCC
on the diffusion in the hadronic stage is negligible in this 
experimental result. 

From Fig.~\ref{fig:2b05}, one can also estimate that 
the value of $T$ is within the range $T\simeq0.04 - 0.06$.
Figures~\ref{fig:4th20}, \ref{fig:40001}, and \ref{fig:40505051}
show that with this value of $T$, the dependence of 
$\langle (Q_{\rm (net)} (\Delta\eta))^4 \rangle_c/\langle Q_{\rm (tot)} (\Delta\eta) \rangle$ on
$\Delta\eta$ for $\Delta\eta/\eta_{\rm tot}\lesssim 0.2$ is 
sensitive to the initial conditions,
such as $[N_{{\rm(tot)}}^2]_{\rm c}$ and $[N_{{\rm(net)}}^n]_{\rm c}$.
In particular, if the fluctuations at the hadronization are well 
suppressed 
the change of the sign of $\langle (Q_{\rm (net)} (\Delta\eta))^4 \rangle_c
/\langle Q_{\rm (tot)} (\Delta\eta) \rangle$ will be observed experimentally
as in Fig.~\ref{fig:4th20}.
In this way, the combination of $\Delta\eta$ dependences of the 
higher order cumulants should be used as an experimental probe
to investigate the primordial thermodynamics at LHC energy.
Of course, the use of the $\Delta\eta$ dependences of the net 
baryon number cumulants \cite{KA} in addition will provide us 
more fruitful information.

Using the above estimate on $T$, the value of the diffusion 
length in the hadronic stage is also estimated as $d(\tau)=0.32 - 0.48$.
Since the diffusion length is directly related to the diffusion 
constant through Eq.~(\ref{eq:d(tau)}), it is possible to make an estimate
on the latter using this relation.
First, we assume that the diffusion constant in the cartesian 
coordinate takes a constant value, $D_H$, in the hadronic medium.
The diffusion constant in Eq.~(\ref{eq:d(tau)}) is the one in the
rapidity coordinate, which is related to $D_H$ as 
\begin{align}
D(\tau)=D_H \tau^{-2}.
\label{eq:D(tau)}
\end{align}
By substituting Eq.~(\ref{eq:D(tau)}) in Eq.~(\ref{eq:d(tau)}) 
and carrying out the $\tau$ integral, we obtain
\begin{align}
D_H=\frac{d(\tau)^2}{2}\left[\frac1{\tau_0}-\frac1{\tau_{\rm fo}}\right]^{-1}.
\label{eq:diff}
\end{align}
Second, 
from the analysis of the dynamical models for LHC energy
\cite{Song:2013tpa}, we estimate the proper times of chemical and kinetic 
freezeout as $\tau_0\simeq8-12$ fm and $\tau_{\rm fo}\simeq20-30$ fm, respectively.
Substituting these values in Eq.~(\ref{eq:diff}), we obtain
\begin{align}
D_H = 0.6-3.5 .
\label{eq:value}
\end{align}
Although this is a rough estimate, 
it is notable that the value is not far from
the ones estimated by combining the lattice simulations  
and the balance function \cite{Ling:2013ksb}.

Finally, we remark the following.
In the above estimate we have implicitly assumed that 
the pseudo-rapidity dependence in Ref.~\cite{ALICE} is identical
with the one of the space-time rapidity $\eta$.
These two rapidities, however, are not identical 
even under the assumption of Bjorken scaling
because the momentum of a particle has a thermal distribution
in the rest frame of the medium.
Owing to this effect, 
the distribution $n(\eta)$ in space-time rapidity at $\tau_{\rm fo}$
is blurred in the experimental measurement by the pseudo-rapidity.
This transformation acts to make the magnitude of 
$\langle (Q_{\rm (net)} (\Delta\eta))^2 \rangle_c$ approach the Poissonian one.
Since the estimate of the diffusion constant in Eq.~(\ref{eq:value}) 
is made without this effect, the value in Eq.~(\ref{eq:value}) 
will become smaller when the effect is appropriately taken into account.
One thus should regard Eq.~(\ref{eq:value}) as the upper limit 
of the estimate of $D_H$.
The effect of the rapidity blurring will be investigated
elsewhere \cite{yohnishi}.
It should also be remembered that the approximate correspondence
between pseudo and space-time rapidities relies on 
the Bjorken scaling.
Although the scaling is qualitatively valid at mid-rapidity 
region for LHC and top-RHIC energies,
it is eventually violated as $\sqrt{s_{\rm NN}}$ becomes smaller.
When fluctuation observables at small $\sqrt{s_{\rm NN}}$ are 
investigated, therefore, one should keep this effects in mind 
as well as the enhancement of the GCC effect 
owing to the decrease of $y_{\rm tot}$ at small $\sqrt{s_{\rm NN}}$.

\section{SUMMARY}

In the present study, we investigated the effect of the GCC on
cumulants of conserved charges in heavy ion collisions by studying the time evolution of
cumulants in a finite volume system with reflecting boundaries in the space-time rapidity space with the diffusion master equation. 
Our result shows that the effect of the GCC appears
in the range of the diffusion length from the boundaries.
This result suggests that the effects of the GCC must 
be investigated dynamically by taking account of the time 
evolution of the system generated in heavy ion collisions. 
By comparing our result with the $\Delta\eta$ dependence of 
net electric charge fluctuation at ALICE \cite{ALICE},
we showed that the effects of the GCC on the diffusion in the 
hadronic medium on cumulants of conserved charges
are almost negligible in the rapidity window available 
at the ALICE detector.

We also emphasized that the $\Delta\eta$ dependence of fluctuations of conserved charges will 
tell us information on the properties and the time evolution of the hot medium generated
in heavy ion collisions, namely the initial charge distribution, the mechanism of hadronization, and the diffusion constant.
Up to now, the second order cumulant of the net electric charge at LHC has been the only fluctuation of conserved charges
whose rapidity window dependence was measured.
If the measurement of the rapidity window dependences of the 
higher order cumulants of net electric charge number, as well
as those of net baryon number, are performed at both RHIC and LHC,
it will make it possible to reveal various aspects of 
the hot medium created by heavy ion collisions.

We thank H.~Song for providing us the data on their dynamical simulation.
M.~S. was supported by Faculty of Science, Osaka University.
This work was supported in part by 
JSPS KAKENHI Grant Numbers 23540307, 25800148, and 26400272.

\end{document}